\documentclass{epl}

\usepackage{epsfig}
\usepackage{graphicx}
\usepackage{dcolumn}
\usepackage{bm}

\title{Observation of particle hole asymmetry and phonon excitations in non-Fermi
liquid systems: A high-resolution photoemission study of ruthenates}

\shorttitle{Particle hole asymmetry and phonons in PES}

\author{Kalobaran Maiti,\footnote{Corresponding author: kbmaiti@tifr.res.in}
Ravi Shankar Singh, and V.R.R. Medicherla}

\institute{Department of Condensed Matter Physics and Materials
Science, Tata Institute of Fundamental Research, Homi Bhabha Road,
Colaba, Mumbai - 400005, INDIA}


\begin{document}

\maketitle

\begin{abstract}

We investigate the temperature evolution of the electronic states in
the vicinity of the Fermi level of a non-Fermi liquid (NFL) system,
CaRuO$_3$ using ultra high-resolution photoemission spectroscopy;
isostructural SrRuO$_3$ exhibiting Fermi liquid behavior despite
similar electron interaction parameters as that of CaRuO$_3$, is
used as a reference. High-energy resolution in this study helps to
reveal particle-hole asymmetry in the excitation spectra of
CaRuO$_3$ in contrast to that in SrRuO$_3$. In addition, we observe
signature of phonon excitations in the photoemission spectra of
CaRuO$_3$ at finite temperatures while these are weak in SrRuO$_3$.

\end{abstract}

Recent observations of various bulk properties in normal phase of
high temperature superconductors \cite{olson}, $d$ and $f$ electrons
at quantum critical points \cite{stewart}, low dimensional systems
\cite{haldane} {\em etc.}, exhibit deviations from Fermi-liquid (FL)
behavior, the most fundamental paradigm in solid-state physics
\cite{pines}. Such non-Fermi liquid (NFL) behavior has often been
attributed to strong electron correlations and/or charge
fractionalizations \cite{denlinger}. Interestingly, CaRuO$_3$, a
3-dimensional orthorhombically distorted perovskite exhibits NFL
behavior at low temperatures, while isostructural SrRuO$_3$ exhibits
Fermi liquid (FL) behavior \cite{klein}. Magnetic measurements
reveal ferromagnetic ground state in SrRuO$_3$ (Curie temperature,
$T_C$ $\sim$ 165~K). However, CaRuO$_3$ does not exhibit long range
order.

The average Ru-O-Ru bond angle is slightly different in these two
compounds (150$^\circ$ in CaRuO$_3$ and 165$^\circ$ in SrRuO$_3$).
It was believed that the decrease in the electron hopping
interaction strength, $t$, due to smaller Ru-O-Ru angle in CaRuO$_3$
leads to an increase in effective electron correlation strength,
$U/W$ ($U$ = electron-electron Coulomb repulsion strength, $t
\propto W$ $\sim$ bandwidth), and hence the NFL behavior appears in
CaRuO$_3$ \cite{klein}. However, a recent study based on band
structure calculation and photoemission spectroscopy \cite{ruth}
shows that the width of various $d$-bands remains almost the same
across the series and $U/W$ is essentially identical in both the
compounds. Thus, significantly different ground state properties
despite similar electron interaction parameters in these systems
pose a challenging question and any comparative study will provide
an ideal opportunity to investigate various fundamental factors
responsible for NFL behavior. With this motivation, we report our
results on the evolution of the electronic structure of CaRuO$_3$
and SrRuO$_3$ as a function of temperature employing photoemission
spectroscopy with ultra high resolutions (1.4~meV for ultraviolet
photoemission spectroscopy and 300~meV for $x$-ray photoemission
spectroscopy). Significantly different temperature evolutions of the
bulk spectra of SrRuO$_3$ and CaRuO$_3$ reveal the signature of
different magnetic ground states in their electronic structure. Very
high resolution enabled us to probe the signature of particle-hole
asymmetry and phonon excitations in the vicinity of the Fermi level,
$\epsilon_F$ in the temperature evolution of photoemission spectra
in CaRuO$_3$ in contrast to that in SrRuO$_3$.

Photoemission measurements were performed on high quality samples
\cite{ruth} at a base pressure of 3$\times$10$^{-11}$~torr using a
SES2002 Gammadata Scienta analyzer and monochromatized photon
sources. The details of the sample preparation and
characterization has been described elsewhere\cite{ruth}. The
sample surface was cleaned by {\it in situ} scraping. The
cleanliness and reproducibility of the spectra were ascertained
after each cycle of scraping.

\begin{figure}
\vspace{-4ex}
 \centerline{\epsfysize=4.5in \epsffile{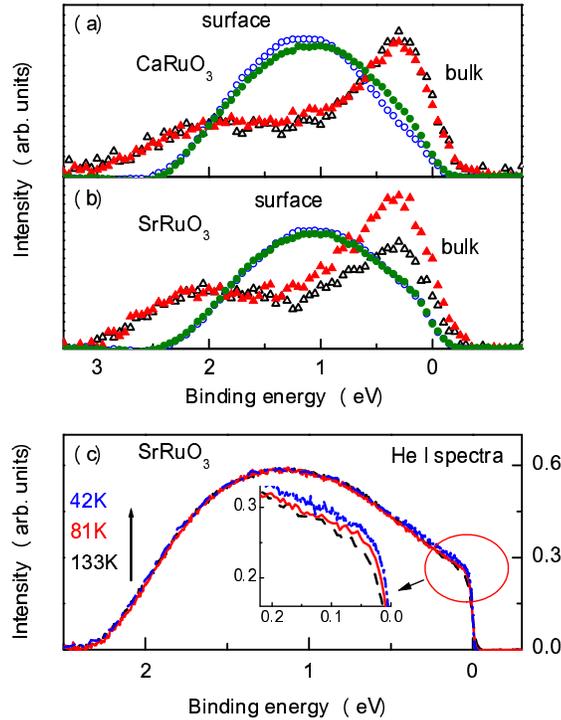}}
\vspace{-6ex}
 \caption{(color online) Surface(circles) and bulk (triangles) spectral
functions of (a) CaRuO$_3$ and (b) SrRuO$_3$. Closed symbols
represent spectra at 300 K and the open ones represent spectra at
48~K and 35~K for CaRuO$_3$ and SrRuO$_3$, respectively. (c) High
resolution He~{\scriptsize I} spectra of SrRuO$_3$. The dashed,
solid and dot-dashed lines represent the spectrum at 133~K, 81 K
and 42~K, respectively. The inset shows the expanded view of the
same spectra near $\epsilon_F$.}
 \vspace{-2ex}
\end{figure}

It is now realized that the surface electronic structure of
transition metal oxides \cite{ruth,lcvo,csvo} is significantly
different from the bulk electronic structures due to different
symmetry, surface reconstruction {\it etc.} that occur at the
surface. Thus, we have extracted the surface and bulk spectral
functions of SrRuO$_3$ and CaRuO$_3$ using high-resolution spectra
at Al~$K\alpha$ (1486.6~eV) and He~{\scriptsize II} (40.8~eV)
radiations across $T_C$ and shown in Fig.~1. Interestingly, the
sharp feature at $\epsilon_F$ in the bulk spectra, known as coherent
feature representing the delocalized electronic states reduce
drastically below $T_C$ in SrRuO$_3$ with respect to the incoherent
feature intensity around 2~eV (signature of correlation induced
localized states). Such reduction in intensity is consistent with
the results from band structure calculations \cite{david,kbmband}
and can be attributed to the finite exchange splitting leading to a
shift of the minority spin contributions above $\epsilon_F$. Thus,
these results suggest that the coherent feature in the 35~K spectrum
primarily appears due to up-spin contributions and that up- and
down-spin Fermi surfaces are significantly different.

Within the ferromagnetic phase, the spectral functions are
relatively insensitive to the change in temperature in the wide
energy scale as evident in the He~{\scriptsize I} (21.2~eV) spectra
in Fig.~1(c). The decrease in temperature leads to a small increase
in intensity near $\epsilon_F$ (signature of the resonant feature)
as evident in the high-resolution spectra shown in an expanded scale
in the inset. It is difficult to realize such evolutions merely from
the Fermi-Dirac distribution function as the energy range of
intensity enhancement is much larger ($>$~200~meV) than expected
($\epsilon_F \pm 2k_BT$) energy range. Various state-of-the-art
calculations based on dynamical mean field theory predicted such
temperature-induced changes in a correlated Fermi liquid system (see
Fig.~39 and corresponding discussions in Ref. \cite{kotliar}). These
thermal effects in the electron density fluctuations were used to
demonstrate the metal-insulator transitions in a correlated electron
system as a function of temperature. While such effects are often
smeared out due to the disorder introducing a dip at $\epsilon_F$,
the high resolution employed in this study helped to reveal this
effect in real systems. In sharp contrast, the bulk spectral
functions in CaRuO$_3$ (see Fig. 1(a)) are essentially identical in
the wide energy scale down to 48~K. High resolution spectra do not
exhibit temperature evolutions as observed in SrRuO$_3$.

Interestingly, the surface spectra shown in Fig. 1(a) and 1(b) for
both CaRuO$_3$ and SrRuO$_3$ exhibit very small intensity at
$\epsilon_F$ at all the temperatures. Resolution broadening of 300
meV of the intense higher binding energy features in the surface
spectra is expected to enhance the intensity at $\epsilon_F$.
Instead, observed small intensities indicate that the spectral
density at $\epsilon_F$ in the surface electronic structure is
negligible. Thus, the intensity at $\epsilon_F$ (coherent feature)
in the He~{\scriptsize I} spectra essentially correspond to the bulk
electronic structure. Such a scenario has indeed been demonstrated
in a similar system, SrVO$_3$ \cite{shen}, where the dispersion and
mass enhancement of the electronic states at $\epsilon_F$ in the
ultra-violet spectra are identical to the bulk of the system.

\begin{figure}
\vspace{-2ex}
 \centerline{\epsfysize=4.5in \epsffile{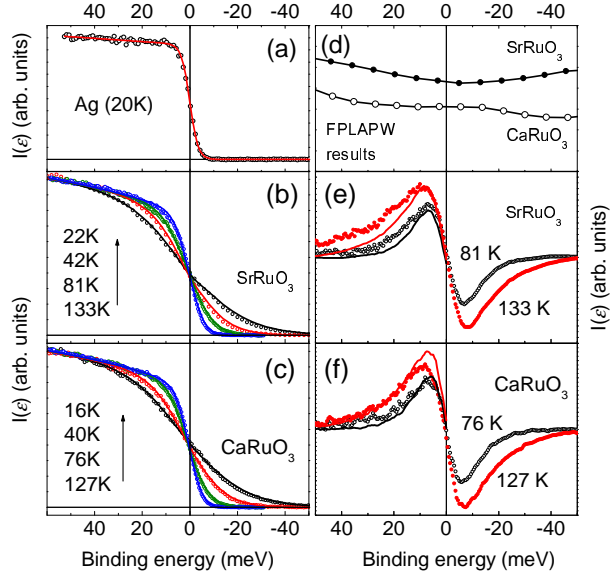}}
\vspace{-22ex}
 \caption{(color online) (a) Valence band spectra of Ag at 20~K near
$\epsilon_F$. The line represents the bare Fermi-Dirac distribution
at 20~K. (b) and (c) show the temperature evolution of the spectral
functions near $\epsilon_F$ for SrRuO$_3$ and CaRuO$_3$,
respectively. (d) Total density of states of SrRuO$_3$ and CaRuO$_3$
from the Full Potential Linearized Augmented Plane Wave
Calculations. (e) and (f) show the difference spectra obtained by
subtracting the spectra at higher temperatures from the lowest
temperature spectra of SrRuO$_3$ and CaRuO$_3$, respectively. The
lines represent the same subtracted spectra above $\epsilon_F$
inverted and superimposed onto the lower energy part.}
 \vspace{-2ex}
\end{figure}

In order to investigate the anomalous behavior in CaRuO$_3$ further,
we probe the spectral evolution close to $\epsilon_F$. The energy
resolution and the Fermi level at each temperature are determined by
the experiments on high purity polycrystalline Ag sample as shown in
Fig.~2(a) for the spectrum at 20 K. Interestingly, the experimental
spectrum (circles) is almost exactly reproduced by bare Fermi-Dirac
distribution function (line) without using any broadening due to the
instrumental resolution. This is again confirmed by comparing the
energy derivative of both the spectral functions across
$\epsilon_F$. A broadening more than 1.5~meV (FWHM) exhibits
deviations from the experimental spectrum and is consistent with the
measured resolution broadening in Xe 5$p$ levels. Since transport
occurs in the low energy scale ($\sim$~meV), {\it it is necessary to
achieve such high resolution to critically investigate these
properties}. Using this {\it state-of-the-art} energy resolution
achieved in this instrument, we investigate the spectral changes in
SrRuO$_3$ and CaRuO$_3$ as shown in Fig.~2(c) and 2(d),
respectively. All the spectra, normalized at the binding energy 100
meV($\gg k_BT$), appear to cross each other at $\epsilon_F$ as
expected in a system of fermions following Fermi-Dirac distribution
function.

The electrical conduction, $\sigma$, of a Fermi-liquid scales as
$T^{-2}$ and can be expressed as $\sigma = {{n(\epsilon)e^2
\tau}\over{m^\star}}$  ($n(\epsilon)$ = carrier density in the
vicinity of $\epsilon_F$, $\tau$ = scattering rate and $m^\star$ =
effective mass). Thus, in addition to the temperature dependence of
$\tau$, the shape of $n(\epsilon)$ and its evolution with
temperature play a significant role in determining the temperature
dependence of electronic conduction. Various recent studies
\cite{florence} suggest that a simple power law dependence of
$n(\epsilon)$ captures most of the physical properties associated to
the electronic states close to $\epsilon_F$. Thus, we simulate the
photoemission response, $I(\epsilon) = \int d\epsilon^\prime\times
n(\epsilon^\prime)\times F(\epsilon^\prime,T)\times
G(\epsilon,\epsilon^\prime,\gamma)$ considering $n(\epsilon) = n_0 +
n_1\mid(\epsilon-\epsilon_F)\mid^\alpha$, $F(\epsilon,T)$ is the
Fermi-Dirac distribution function and $G$ is the Gaussian
broadening. We find that $\alpha$ = 0.5$\pm$0.05 provides a good
description of the spectra for both CaRuO$_3$ and SrRuO$_3$ for $BE
> 2k_BT$ suggesting an influence of disorder in the electronic structure as
also observed in other oxides \cite{altschuler,ddprl}.
Interestingly, the full width at half maximum (FWHM) of the Gaussian
needed to simulate the experimental spectra is significantly large
compared to 1.4~meV expected from the instrument resolution
function. Most notable change is observed in the case of CaRuO$_3$.
While FWHM in SrRuO$_3$ is always $<$~8~meV, the values in CaRuO$_3$
are 14.5, 11.0, 7.8 and 7.5~meV at the temperatures of 127~K, 76~K,
40~K and 16~K respectively. Such significant temperature dependence
of FWHM suggests influence of electron-phonon coupling in the
electronic structure of CaRuO$_3$ \cite{vallaprl}.

In addition, the spectral weight transfer appears to be different in
the two systems; e.g. the spectral weight transfer from 0-20~meV
binding energy range to the energy region above $\epsilon_F$ for a
change in temperature from 22~K to 42~K in SrRuO$_3$ is
significantly larger compared to that observed for similar
temperature change (16~K to 40~K) in CaRuO$_3$. A careful look at
the spectra reveals two major differences between the two systems.
Firstly, the intensities very close to $\epsilon_F$
$(\mid\epsilon-\epsilon_F\mid < k_BT)$ exhibit an anomaly between
the fit and experimental spectra. In order to bring out this point,
we subtract the spectral functions at different temperatures from
the lowest temperature spectra and show them in Fig.~2(e) and 2(f).
The lines represent the spectral differences above $\epsilon_F$,
superimposed after inversion onto the difference spectra below
$\epsilon_F$. In the case of SrRuO$_3$, the difference spectra are
almost symmetric with respect to $\epsilon_F$ for
$(\mid\epsilon-\epsilon_F\mid < k_BT)$ as expected for a
Fermi-liquid system. The differences observed around 10 - 50~meV
binding energies may be attributed to the temperature induced
gradual population of the coherent feature as shown in the inset of
Fig.~1 \cite{kotliar}. In sharp contrast, the spectra in CaRuO$_3$
reveal unusual evolution with temperature; the spectral weight
transferred above $\epsilon_F$ is larger than the reduction in
intensity below $\epsilon_F$. This is clearly in contrast to the
expected trend based on band structure results shown in Fig.~2(d),
which shows higher intensity above $\epsilon_F$ in SrRuO$_3$
suggesting a larger spectral weight transfer than that in CaRuO$_3$.
{\it Increase in temperature populates the hole states, which could
be probed efficiently using high energy resolution.} Thus, the
anomaly observed in Fig.~2(f) clearly reveals the asymmetric
excitations between electrons and holes (particle-hole asymmetry)
\cite{anderson} in CaRuO$_3$.

\begin{figure}
\vspace{-4ex}
 \centerline{\epsfysize=4.0in \epsffile{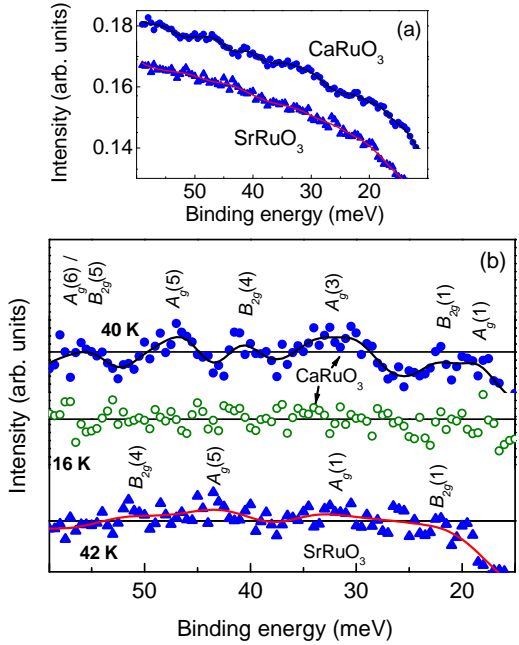}}
\vspace{-6ex}
 \caption{(color online) (a) The expanded raw spectra of CaRuO$_3$ and
SrRuO$_3$ at 40~K in the binding energy range 20-60~meV. (b)
Spectral features obtained by subtracting the simulated spectral
function shown in Fig. (2) at different temperatures. $A_g$(1) =
RuO$_6$ $z$-rotation, Ca; $A_g$(3) = RuO$_6$ rotation/stretching;
$A_g$(5)= Ca($y$), RuO$_6$ stretching/bending; $A_g$(6) = O-Ru-O
bendings; $B_{2g}$(1) = Ca($x$), RuO$_6$ breathing; $B_{2g}$(4) =
O-O stretching; $B_{2g}$(5) = O-Ru-O bendings as per the
assignments in Refs. \cite{kolev}.}
 \vspace{-2ex}
\end{figure}

In addition, the spectral function of CaRuO$_3$ exhibits pronounced
oscillations compared to that in SrRuO$_3$ at 40~K as shown in
Fig.~3(a). This can clearly be seen in Fig.~3(b), where we show the
spectral functions after subtracting the fits shown by lines in
Fig.~2(c) and 2(d). The oscillations are weak in SrRuO$_3$ and
appear at different binding energies; the experimental data for both
CaRuO$_3$ and SrRuO$_3$ were collected in the same set up with both
the samples mounted together on the same sample holder. {\it The
observation of such oscillations is unusual and has only been
possible for the first time in the photoemission spectroscopy due to
the very high-energy resolution employed in this study.} It is to
note here that such oscillations in density of states was observed
in point contact\cite{yanson}, tunneling studies\cite{geerk}, and
was attributed to the electron-phonon coupling
effect\cite{carbotte}. We, thus, compare the energy positions of the
peaks with the phonon response observed in the Raman spectra in
various recent studies \cite{kolev}. Interestingly, all the signals
observed here correspond to rotation/stretching/bending of RuO$_6$
octahedra, and could be assigned based on the Raman spectra and
subsequent lattice dynamics calculations.

Interestingly, the distinct peaks at 40~K in CaRuO$_3$ spectrum
become invisible at 16~K. The reduction in temperature leads to a
decreased degree of electron-phonon coupling. However, the
disappearance of the features at 16 K compared to the observation at
40 K may need attention. While it may be likely that the signature
of coupling is below the noise level of the measurements, further
studies are required (both theoretical and experimental) to probe
this effect. We believe that these results would help to initiate
investigations in this direction.

In SrRuO$_3$, the electron-phonon coupling appears to be weaker and
total 4 modes corresponding to breathing modes of RuO$_6$ octahedra
and O-Ru-O bending could be observed in the energy range studied.
The spectrum in CaRuO$_3$, however, exhibit many additional phonon
modes. It is to note here that all the additional modes in CaRuO$_3$
correspond to RuO$_6$ rotation/stretching, different O-Ru-O bending.
This indicates that a smaller Ru-O-Ru angle in CaRuO$_3$ is more
susceptible for strong scattering of electrons by the lattice.
Interaction of electrons and lattice vibrations is known to be
crucial to determine various exotic material properties such as
colossal magnetoresistance, pseudogap phase in high-temperature
superconductors \cite{millis,littlewood}. This study indicates that
the role of electron-phonon coupling in various novel material
properties needs critical considerations.

NFL behavior is often found experimentally in a phase in the
proximity of quantum critical point, where the NFL ground state is
related to magnetic instability \cite{stewart}. Evidence of the
proximity of such quantum critical behavior has indeed been observed
in the recent studies in high-temperature superconductors
\cite{vallascience}. Various studies suggest that existence of
low-dimensionality in these systems leads to charge
fractionalization and hence NFL behavior manifests as has been shown
in one-dimensional systems possessing decoupled charge and spin
excitations \cite{vallanature,orgadprl}. The signature of
particle-hole asymmetry and phonon excitations in the spectral
functions of CaRuO$_3$ observed in this study provide evidences to
consider additional parameters in the study of NFL behavior.

The authors acknowledge useful discussions with Prof. P.W. Anderson,
Princeton, USA.

\end{document}